\documentclass{wg119}
\normallatexbib
\usepackage {epsfig}
\usepackage{graphicx}
\usepackage{booktabs}
\usepackage[hyphens]{url}

\begin{document}

\articletitle{Network Intell: Enabling the Non-Expert Analysis of Large Volumes of Intercepted Network Traffic}

\chaptitlerunninghead{van de Wiel, Scanlon \& Le-Khac}

\author{Erwin van de Wiel, Mark Scanlon and Nhien-An Le-Khac}

\begin{abstract}
In criminal investigations, telecommunication wiretaps have become a common technique used by law enforcement. While phone-based wiretapping is well documented and the procedure for their execution are well known, the same cannot be said for Internet taps. Lawfully intercepted network traffic often contains a lot of encrypted traffic making it increasingly difficult to find useful information inside the traffic captured. The advent of Internet-of-Things further complicates the process for non-technical investigators. The current level of complexity of intercepted network traffic is close to a point where data cannot be analysed without supervision of a digital investigator with advanced network knowledge. Current investigations focus on analysing all traffic in a chronological manner and are predominately conducted on the data contents of the intercepted traffic. This approach often becomes overly arduous when the amount of data to be analysed becomes very large. In this paper, we propose a novel approach to analyse large amounts of intercepted network traffic based on network metadata. Our approach significantly reduces the duration of the analysis and also produces an insight view of analysing results for the non-technical investigator. We also test our approach with a large sample of network traffic data.
\end{abstract}


\section{Introduction}

Lawful interception is a method that is used by the police force in some countries in almost all middle-to high-level criminal investigations. The investigation of intercepted/wiretapped data can be classified into two objects, voice and network data~\cite{vsl:Schut}. This paper will focus on intercepted network data, which is a necessary source of evidence for a range of crimes, e.g., child abuse material distribution~\cite{vsl:Yasinsac}, cloud hosted services~\cite{vsl:Schut}, industrial espionage~\cite{vsl:Nelson}, dead drops~\cite{vsl:Nelson, vsl:bysync}, malicious software distribution~\cite{vsl:bysync}, instant messaging~\cite{vsl:Walnycky}, piracy~\cite{vsl:hannaway}, illegal content distribution~\cite{vsl:bysync}, etc. A problem arises for non-technical detectives as it is very difficult to interpret the network data. Analysis of the data that is generated with such investigations can also be very labour-intensive. The rules for placing a telephone or IP tap constitute a special power of investigation. In the Netherlands, for example, Dutch law has provided this power with a legal base that is called The Special Investigative Powers Act (SIPA or Bijzondere Opsporings Bevoegdheid, BOB in Dutch). Dutch Telecom Law stipulates that every provider must have equipment for data interception for law enforcement. It is also allowed to intercept data of locations without a service provider. This can be used for example to intercept data from wireless access points.

While ideally, all digital forensic evidence would be analysed by expert analysts, the reality is that there are simply too many cases requiring expert analysis and the backlog of cases is often too large~\cite{vsl:Quick}. As a result, non-expert evidence processing has become a necessity in many law enforcement agencies around the world~\cite{vsl:battling}. Analysis of intercepted data from an IP tap is often conducted by non-technical investigators. These investigators will analyse the data in a chronological manner. One example scenario could be when the investigator needs to analyse data for a period of four weeks. The investigator will start at day one of the IP tap data from the analysis software. The analysis software presents all collected data for that day starting from 0:00 to 23:59. An example focus might be web traffic using HTTP. Each web visit is displayed in a list view. The user can select a row and the contents of the website will be displayed. The only option for the investigator would be to look at every web visit sequentially. Hence, it takes a lot of time to analyse an IP tap for a period of four weeks. The analysis software has the option to filter on text or protocol like HTTP or FTP, but the investigator must know what text to filter beforehand. One method to create a good filter would be to analyse every day of the IP tap and create text filters based on that knowledge. All intercepted traffic will appear as coming from a single internet address and the investigating officer needs to identify what source is causing the network traffic from the intercepted data. Looking at the current amount of internet traffic an average user is producing; the volume requiring analysis is almost impossible to keep on top of and is set to further increase with the additional viability of cloud network traffic to digital investigations~\cite{vsl:Lillard, vsl:cloud}.

In this paper, we propose a novel approach that can produce usable results out of a big stream of intercepted data from an IP tap. Instead of analysing the data chronologically, the program identifies what lies behind the intercepted internet connection and produces an overview of the found information. This result can be shown directly after analysis and non-technical investigators can interpret the data. This result set can be used in the investigation as tactical information or it can be used to investigate the data at a deeper level with the help of a digital investigator. This paper also describes the development of a proof-of-concept program called ``Network Intel'' that will introduce a novel way to analyse intercepted data from an IP tap. This program does not focus on the data contents, but analyses the metadata of the different protocols that can be found inside the data. The goal is to reduce the time consumed in analysing an IP tap. We can also use this method as a quick way to retrieve intelligence information. This intelligence information consists of a clear view what can be seen behind an IP tap. For example, when the investigator knows that a mobile phone device is present, he can use this information in the investigation. This information could prove valuable for the investigation, because investigative decisions can be made based on that intelligence. We also test our approach with varied large size of network traffic data to evaluate its performance.

\section{Background} 

Several tools have been developed for the analysis of intercepted network data. Digital investigators require specialised knowledge to use the array of tools needed~\cite{vsl:Casey}. The majority of digital forensic network investigation tools available are only usable by people with highly-specialised knowledge. 

Current tools can be divided in two groups. The first group includes tools that can be used only by people with a high level of network knowledge like network administrators or digital investigators with advanced network protocol knowledge. These tools are used to analyse network traffic at the packet level and can be used to create extractions of data that can be used in deeper analysis of the capture. The second group consists of specialised network forensic tools. These are the well-known tools that are used to analyse captured network traffic. Although these tools are not as complex as the first group, they still need to be operated and interpreted by digital investigators with specialised network knowledge. Most of these tools are freeware or a professional version can be bought at a price of around 500-1000 US dollars. All tools operate on the well-known packet capture format PCAP.

\subsection{Network Forensic Analysis Tools (NFATs)} 
NetworkMiner is a tool that was created and released by Erik Hjelmvik in 2007~\cite{vsl:Hjelmvik}. This tool splits and extracts different items inside network traffic. Although it creates a good overview the tool still needs to be operated by someone with network skills. The latest version of NetworkMiner has the option to distinguish between hosts, but this is not usable when a lawful intercepted network connection is analysed. The captured traffic then seems to come from a single IP address (the tapped IP address). NetworkMiner does not scale well with significant data size. For reference, the analysis of a 1.25 GB data capture takes over 21 minutes. Network Miner will parse all information and the result can be analysed by scrolling the list view. This tool is possibly too technical to be used by non-expert investigators.

Xplico is a tool that extracts application data from a network capture~\cite{vsl:Costa}. Network traffic is broken down to protocol level and for each protocol all metadata is extracted. The Xplico tool is usable to analyse larger datasets of network captures. It will parse all information and the result can be analysed by clicking through the web interface. Although the tool is easier to operate, an investigator still needs to analyse/review all the data. Again, the problem that it is not possible to identify different devices behind the captured network point without knowledge to interpret the data.

Wireshark, originally named Ethereal, is perhaps the most well-known network packet analyser~\cite{vsl:Wireshark}. Traffic is broken down to the packet level and broken or half packets can still be analysed. Wireshark also has the option to reconstruct network flows and data can be analysed with the use of advanced filtering. Although the tool is very usable for deep packet inspection, it is not easily usable by non-technical investigators. It also has quite slow performance on filtering and running modules. Each filter and/or module needs to be run on the entire dataset. Creation of filters can be also very slow, because each filter needs to be tested on the full database again. Only single capture files are supported. There is the possibility to connect different captures together with third-party tools. Using this tool on intercepted network traffic of residential/corporate users consumes a significant amount of time to analyse large volumes of information.

\subsection{Approaches for Analysing Metadata from Intercepted Network Traffic}

There are several tools available that examine and/or extract metadata from capture files. One example is the \textit{p0f} tool~\cite{vsl:Zalewski}. This tool is using a passive traffic fingerprinting mechanism to identify players behind TCP/IP communications. To get the best results, the tool must be run on the intercepted network. For lawful interception, this is almost never possible due to the fact that the actual capture will take place at the Internet provider. p0f will only identify or fingerprint traffic coming from the tapped IP address (most of the time a single router or gateway). Another tool that also focuses on metadata from capture files is \textit{Justniffer}~\cite{vsl:Justniffer}. This tool is targeted at request and response protocol information from different protocols. Justniffer can export information in log format and it is also possible to use python scripts in this tool. 

\section{Introducing Network Intell}
In this section, we present first of all the requirement of a new approach of analysing of huge amount of intercepted network traffic. Our novel approach is developed as a proof-of-concept tool named Network Intell.

\subsection{Requirements}

The proposed approach does not of result in a one-size-fits-all solution, but it will greatly help in the analysis of traffic to discover evidence and successfully investigate a criminal case. Another requirement for our approach is optimised analysis speed. Casey proposed a number of requirements for tools used to process network traffic~\cite{vsl:Casey}. For our approach, it meets the following requirements:

\begin{enumerate}
\item Support \textit{Tcpdump} format: This requirement will be met since the parsing of the network traffic will be done with the help of the \textit{WinPcap} library. The program will support \textit{pcap} type traffic.
\item Reliable protocol identification: This requirement will be met and where possible the use of deep packet inspection will be used to identify different protocols.
\item Data reduction: Actual contents of rebuild packet streams will not be stored. The program will focus on metadata and will only store packet information. This will also increase the automated analysis speed and the need for storage will be reduced.
\item Keyword search capabilities: The program is developed as a tool to identify devices and usage behind an internet connection. Selection of these items will be based on custom made queries.  So it will be possible to search for metadata items based on keywords. Since data contents are not stored there will not be a possibility to search inside data.
\item Read-only during examination: Traffic is parsed inside a SQLite database. Since this database is a result of the original capture it is possible to access the data, but there is no functionality in the program that would need this requirement. After parsing the data no more changes are made to the database of analysed information.
\end{enumerate}

The program must be usable by technical and non-technical personnel alike. This results in the requirement that it should be operable by digital investigators and/or non-technical investigators. The digital investigator can create and change rules to identify devices and objects behind an intercepted network point. The non-technical investigator can simply run the program and report the results of the analysis for evidence in the criminal case or as additional information for further analysis by a digital investigator. Each user is allowed to edit the rulesets.

\begin{figure}[!ht]
\centering
\includegraphics[keepaspectratio=true, scale=1, width=\textwidth]{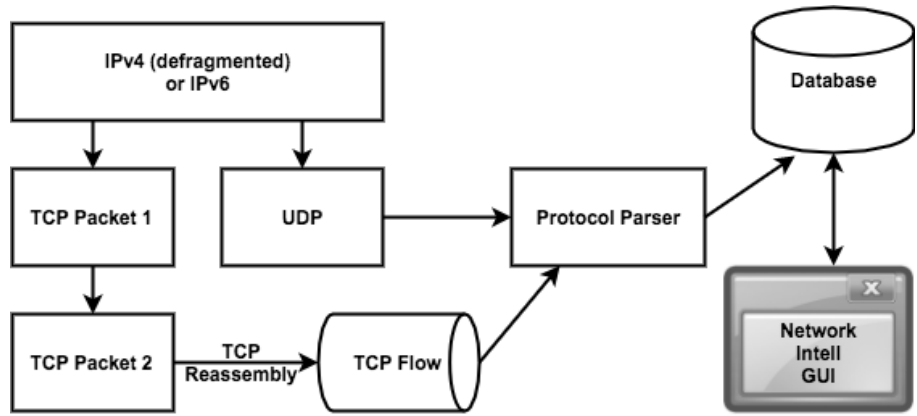}
\caption{Overview of Network Intell Architecture.}
\label{vslfig:one}
\end{figure}

\subsection{Design}

Figure~\ref{vslfig:one} shows the architecture of Network Intell that consists of three main components: Network Parser, Fragmented Traffic Reassembler and Protocol Analysis/Parsing.

\subsubsection{Network Parser}
The premise behind the program is to split protocol data in usable metadata. This metadata will then be stored in a SQLite database and can be queried later on for devices and objects behind the intercept network address. The technical requirements will define which traffic will be used as a parsing object. From that traffic, the most useful protocols will be selected for parsing. Each parsed protocol will be stored in its own database table.

WinPcap~\cite{vsl:WinPcap} has the option to intercept live network traffic or to open stored packet captures (\textit{pcap}). Our tool is developed to only operate on stored packet captures. As stated before the interception of network packets is done at a service provider. The function in WinPcap that is used to open offline stored packet captures is \textit{pcap\_open\_offline}. By opening a packet capture this way it is possible to loop all stored packets. 

The first requirement is set to only include Ethernet packets from the network access layer. The next requirement will be to only parse Ethernet packets with an IPv4 or IPv6 header from the internet layer. As of June 2016, 10.88\% of machines accessing Google used IPv6~\cite{vsl:Nicolls} and this has grown to 18.09\% in January 2018~\cite{vsl:Google}. The parser now selects Ethernet packets with an IPv4 of IPv6 header and a protocol of TCP, UDP or ICMP. At this stage, the ICMP packets are also logged. The final requirement is the application layer. This layer is the most important, due to the final results that need to be stored in the database, which are based on the selection of these protocols.

\subsubsection{Reassembling the Fragmented Network Traffic}
The previous sub-section described the packets that should be used for parsing. The most complex part of network analysis is the reassembly of network packets. Packet reassembly occurs on two layers. The IP layer and the TCP layer. The following solution is used in Network Intell. IP fragmentation can be caused by IPv4 routers that fragment the IPv4 packets when packets are transferred to other networks~\cite{vsl:Olifer}. Reassembly of these fragmented packets is done by the receiving endpoint. Since IPv4 packets do not necessarily to arrive in order, it is a complex process to reassemble the packets in the right order without errors. Reassembly of fragmented packets is therefore done with the use of RFC 815~\cite{vsl:Reassembly} that describes IP datagram reassembly algorithms. This code was implemented in the \textit{IPTraf} tool~\cite{vsl:IPTraf} and is used in Network Intell. Several attack vectors exist against the manner with which IP fragmentation is handled. However, protection against IP fragmentation attacks is beyond the scope of this paper. IPv4 fragmentation is not seen often. IPv6 routers do not support fragmentation of IPv6 packets.

TCP reassembly is the second step for rebuilding network streams and data. Network analysis tools rely on good TCP reassembly to rebuild traffic. The following illustration will show how TCP assembly and reassembly works.

Rebuilding TCP packets will result in streams or are also called flows. A lot of research has been done on TCP reassembly. Extensive research on TCP reassembly was conducted by Wagener et al.~\cite{vsl:Wagener}. Problems that occur at IPv4 reassembly also occur in TCP reassembly. Exploits can be used to bypass the reassembly process and can even crash the software. Protection against TCP reassembly attacks is not in the scope of this paper. It could be a future test scenario on the software. The proof-of-concept tool is using part of the source code from the TCP reassembly tool \textit{tcpick}~\cite{vsl:tcpick}. There was also a requirement added that every TCP connection must include a complete three-way handshake. The TCP three-way handshake are an exchange of packets before a TCP connection is established. The following illustration will show this exchange of packets before a TCP connection is established.

Three packets are exchanged between client and server with the first SYN identified packet being the most important packet. Ack and seq numbers are used in the reassembly process. The sender of the SYN packet will be the initiator of the connection and can be identified as the ``TCP flow from''. Without a SYN packet it would become very complex to identify who started a connection since packets do not have to arrive in the same order as they are send. Therefore the full three-way handshake is needed before a connection will be parsed to a higher-level protocol. 

The technique used in Network Intell is based on the 4-tuple mechanism: Source IP address, Destination IP address, Source Port, and Destination port. Each reassembled connection has to start with a SYN. This is needed to identify who started the connection so what side is client and what side is server. It is possible to create flows that do not have a starting SYN, but that is a complex technique and can produce false results. It is possible that sending and receiving endpoints get switched and analysis on traffic makes the wrong assumption. An example would be when a HTTP request is made from outside directed to the intercepted network point. Without a SYN packet this could be identified as traffic coming from the inside to the outside of the intercepted network point. As mentioned Network Intell is using the 4-tuple way to identify flows for reassembly. Each flow will be rebuilt in memory and the metadata from this flow will be stored in the database. This means what is the client IP what is the server IP how long was the flow active and how much traffic was intercepted. Flows remain active till a RST of FIN is received on the TCP flow. The flow information metadata is exported to the database and the flow is removed from memory. Network Intell also has the option to export the full flow. This can be useful to analyse specific connections and data transfers. This option is currently disabled, because disk I/O will slow down the analysis. When flows do not receive end packets RST or FIN then that flow needs to be shut down after a period of time. This possibility is built into Network Intell, but this needs more testing to find the correct timing on flow termination. Network Intell is working on captured traffic so real-time analysis on live traffic is not an issue. When IP packets are defragmented and TCP packets are reassembled to stream or flows, they can be parsed by the protocol analyser.

\subsubsection{Protocol Analysis and Parsing}
At this stage of the process the network traffic is ready for analysis by the protocol parser. The first step will be the identification of the contents of the network stream. Casey advised not to filter traffic based on protocol because of the risk that other traffic can be tunnelled through a protocol like HTTP~\cite{vsl:Casey}. This advice is also used to not filter network traffic based on port number. Although the port number will be exported in log files the actual contents of the reassembled packets will be used to identify the protocol used at application layer. The only exception is the DNS protocol running on port 53. This protocol is analysed based on the port number that is used. Filtering on the contents of the TCP streams is also known as Deep Packet Inspection (DPI). DPI can be a very useful approach to detect hidden protocols inside network data. DPI can also be a complex process, because it can cause false hits based on the keywords used to identify different protocols. DPI is a process that should be evaluated often and adjusted there were needed. This needs to be done by a skilled network forensic investigator. The most important part of the software will be at this point. The different application layer protocols will be analysed and the most important metadata will be stored in a SQLite database. The following information will be stored.

The technique used in Network Intell to do DPI is based on the search for specific keywords in the header. First it looks for protocol specific details like a HTTP header ends with ``\textbackslash r\textbackslash n\textbackslash r\textbackslash n'' then it searches for specific keywords that are bound to the used protocol like ``GET'' or ``POST''. Based on the result the reconstructed packets are identified by that protocol. This technique is taken from the approach that Xplico is using to identify specific protocols. Network Intell is using DPI to identify HTTP, FTP, SMTP and POP3. When these protocols are running over other ports then the default ports, the protocol will still be detected because of the DPI technique used.

\section{Implementation}
\begin{figure}[!ht]
\centering
\includegraphics[keepaspectratio=true, scale=1, width=\textwidth, trim={0.1cm 0cm 9cm 0cm}, clip=true]{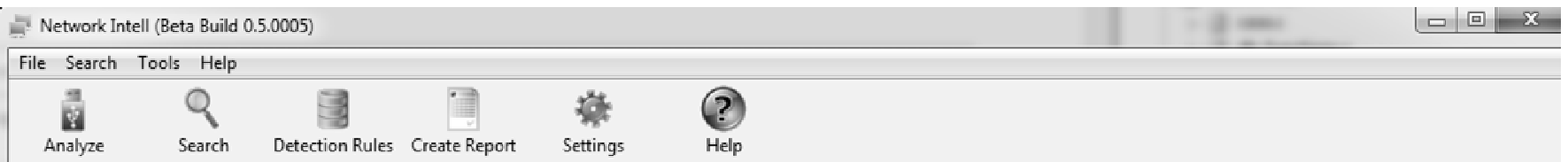}
\caption{Network Intell Main Menu}
\label{vslfig:two}
\end{figure}

Network Intell is a proof-of-concept tool of our approach that can analyse and parse intercepted network traffic. Since speed of the analysis is a requirement for the program the programming language Win32 C is chosen as the programming language. The program will run under Microsoft Windows and the development environment is called Pelles~C~\cite{vsl:Pelles}. Pelles~C is development environment under Microsoft Windows to create 32/64-bit programs in the C language. In Network Intell there are many logs. Each analysis and network packet can be exported to a log file. The program also has the possibility to export each TCP stream/flow in a binary file. This way research can be done on unknown network traffic or content can be verified for consistency. Figure~\ref{vslfig:two} shows the main functionalities of Network Intell including Analyse, Search, Detection Rules, Creating Report, Settings and Help.

\begin{figure}[!ht]
\centering
\includegraphics[keepaspectratio=true, scale=1]{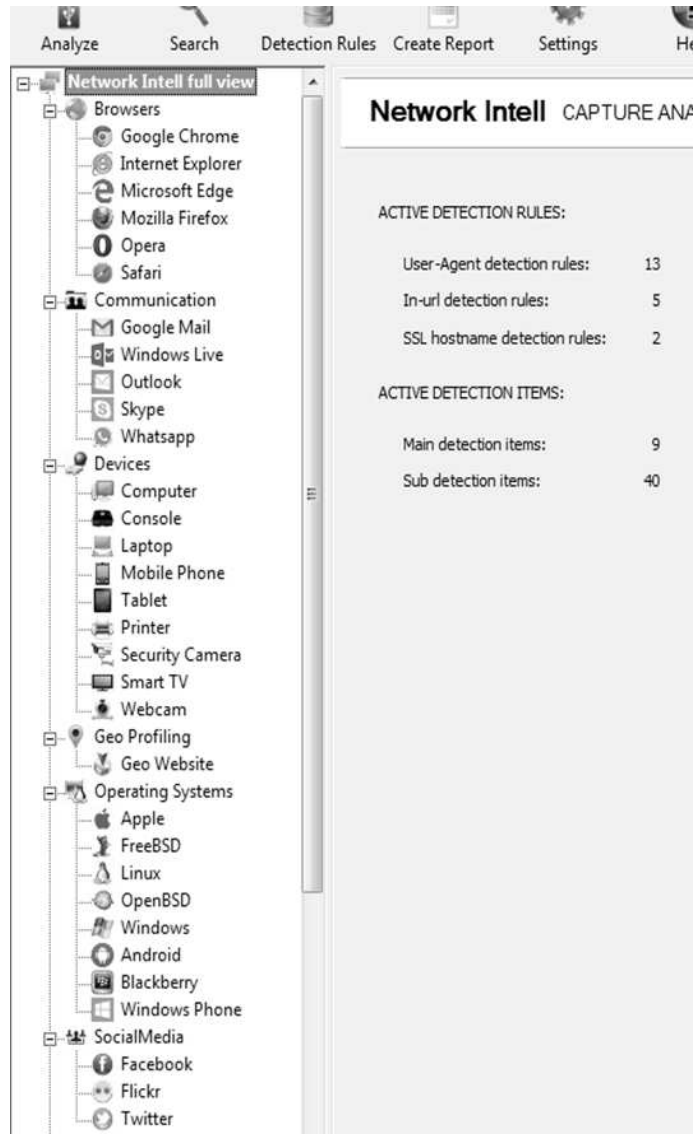}
\caption{Network Intell Full Analysis Options}
\label{vslfig:three}
\end{figure}

Analyse is one of the main functions. This function is the actual start of the analysis. If the investigator knows what capture files need to be investigated then this window can be operated by a non-technical investigator. The start simply asks for an analysis name and the user can add network capture files of the pcap type. When files are added the MD5 and SHA1 hashes are calculated. Also, a count on number of packets is done when loading a capture. The pcap format does not save pack counts in the header. The only safe way to count packets is by opening the capture file and loop all packets with a counter running. The result of the analysis should be a tree view with an overview of analysed items. Since all metadata is stored in a SQLite database, the user is very flexible to include new items, as can be seen in Figure~\ref{vslfig:three}.

\begin{figure}[!ht]
\centering
\includegraphics[keepaspectratio=true, scale=1, width=\textwidth,trim={0.76cm 0.72cm 0.78cm 0.93cm}, clip=true]{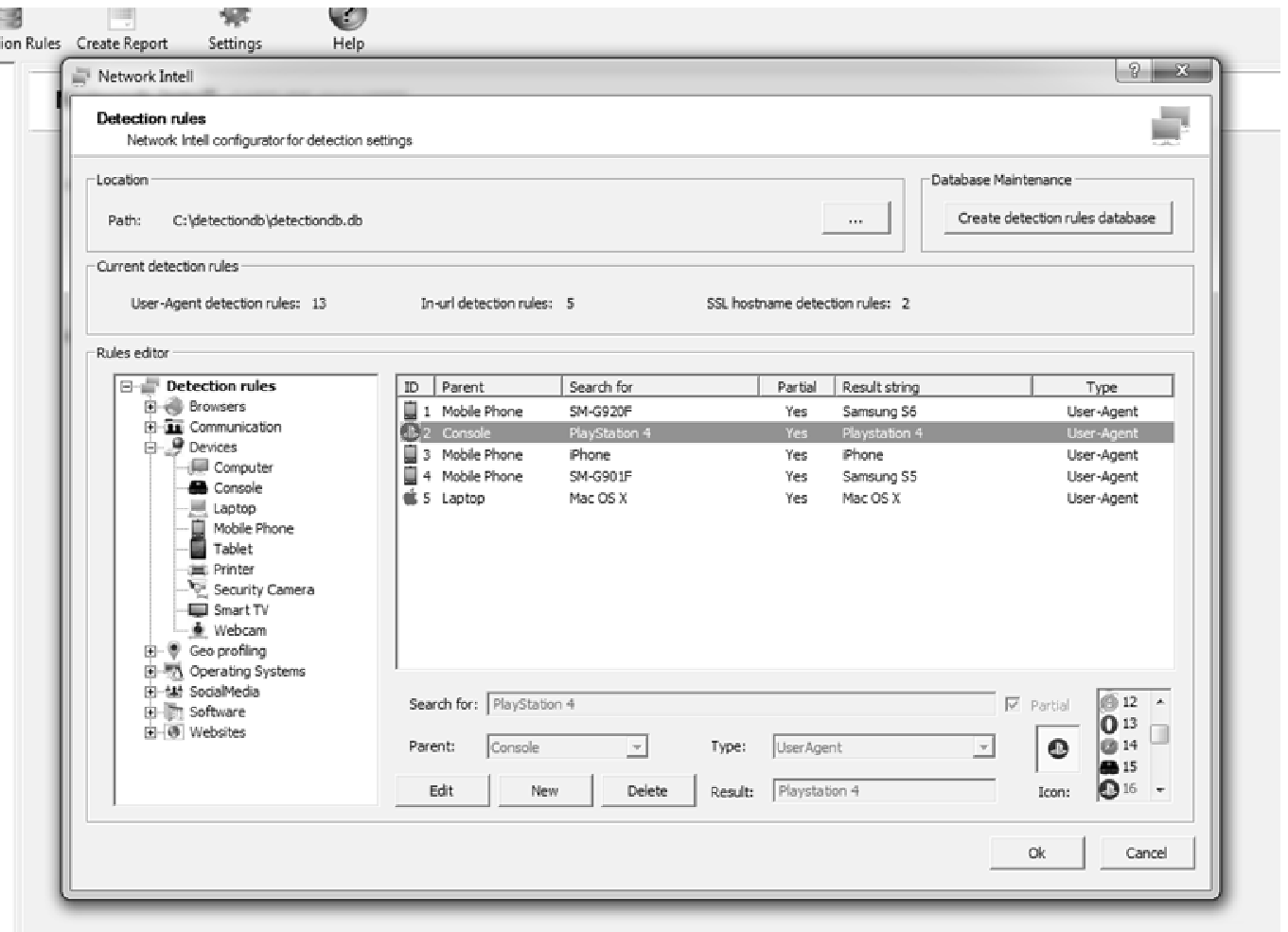}
\caption{Network Intell Detection Rules Editor}
\label{vslfig:four}
\end{figure}

Detection Rule is another important function in Network Intell. A detection rule editor was created in Network Intell for the creation of specific detection rules. This editor is developed with ease-of-use in mind. This editor follows the rules that are mentioned in this section. As a result, a detection rule type has to be selected, result name has to be entered, the parent needs to be selected and of course the search query has to be entered with the option to search partial or the hit needs to be fully equal to the search query, as can be seen in Figure~\ref{vslfig:four}.

\section{Evaluation}

This section will describe what has been the outcome of our approach to solve the research problem. A lot of intercepted network traffic was tested on Network Intell. These were mostly open source capture files or capture files from our home network. The following test method was used: 

\begin{itemize}
\item Each protocol is tested with a protocol specific pcap file.
\item Results from Network Intell were compared with other network forensics tools like Wireshark. Only differences were seen at TCP stream/flow level. Wireshark also identify streams without a starting SYN packet. Network Intell does not identify those streams and need a SYN packet to identify a TCP stream/flow.
\item Personal network traffic was captured at router level to mimic lawful intercepted network traffic. Most errors were detected with this test method, because a lot of traffic can contain content that is not handled by Network Intell. An example was the detection of secure DNS traffic that caused an error in the parsing code. 
\item No real-world case has been tested yet against Network Intell. The tool must be extended with usable functionality. The tool is developed with a requirement that it can be used by anyone. No complex configurations will be needed to use the final tool.
\end{itemize}



Speed tests were conducted to investigate how long it would take to analyse an average network tap. The result is that Network Intell can analyse and export about 1 GB (Gigabyte) per minute depending on the equipment that the software is running on. This number is without the option to export every TCP stream/flow, because that will increase the analysis time a lot. For each TCP stream/flow a new file needs to be created and that will take a long time depending on the number of streams/flows in the captured traffic.  Timers were used in the program to calculate the overall execution time. When the execution data was analysed, it was noticed that there was a lot of disk activity. This was caused by the extensive logging for error checking and program flow analysis. This increased the execution time significantly. As a result, the first measurements were taken with logging enabled. 1GB of test data was used with logging enabled and was measured as taking 63 second on our standard workstation. 
After disabling the logging options, a second test was run. This time, the same 1GB of test data took only 35 seconds to process using Network Intell. 

The result is a database full of protocol specific information and can also be used for further analysis with a SQLite viewer. A query against the SQLite capture database almost runs instantly. The only limitation will be when the database will reach the limitation of the maximum size of a SQLite database. That is in excess of 140 terabytes. No performance testing was done on the analysis since it takes less than a second to produce the resulting tree view. The following image are the statistics that will be presented when the analysis of a pcap file is finished. This info can be used by a digital investigator to check if any errors occurred and how many different streams were analysed.

The adopted approach was to develop a program that can produce usable results out of a big stream of intercepted data from an IP tap. Instead of analysing the data chronologically the program will identify what lies behind the intercepted internet connection and produce an overview of the found information. Besides, the tool is easy to use. The end result is a program that can be operated by almost a single button action. The tool is moreover fast. The end result is a program that is written in c and it will analyse data with a speed of around 1 GB per minute with logging enabled and it takes around 34 seconds for 1GB to be analysed without logging enabled. Looking at other available tools that is a good analysis speed. Our tool is also able to handle Big Data amounts. Tests have been done with capture files of at least 1 GB in size. A lot of network investigation tools cannot handle files of that size or become very slow after loading such amount of data.

\section{Conclusion}
In this paper, we presented a new approach of analysing of huge amount of intercepted network traffic and we developed our approach as a proof-of-concept tool for both technical and non-technical investigators. Our tool is developed that can analyse intercepted network traffic (pcap format) at a speed of around 20 seconds per 200 MB. The analysed information is stored in a SQLite database and can be queried with custom-made statements to create an overview of detected devices and statistics about the analysed traffic. A technical background is not needed to use the program and it can be used with almost a single click action.

\subsection{Future Work}
As Network Intell is a functional, proof-of-concept tool, there are numerous future enhancements possible. In the current version, the detection is based user entered detection rules. A future avenue for development would be to use the Fingerbank database~\cite{vsl:Spooren}, accessible at \url{http://www.fingerbank.org}, or similar. This is a device database with MAC address and User-Agent device detection. Analysis based on the Fingerbank database learns that there are a lot of errors and doubles in the database. Also, from the lawful interception point of view, it is ineffective to use full user agent strings when browsing software is updated almost daily. It is better to create detection rules based on the browser name, than a full user agent string with version numbers. 

\normallatexbib


\begin{chapthebibliography}{99}






\bibitem{vsl:Schut} H. Schut, M. Scanlon, J. Farina, and N-A. Le-Khac, Towards the Forensic Identification and Investigation of Cloud Hosted Servers through Non-Invasive Wiretaps, {\em In Proceedings of 10th International Conference on Availability, Reliability and Security (ARES)}, pp. 249-257. IEEE, 2015.

\bibitem{vsl:Nicolls} V. Nicolls, N-A. Le-Khac, L. Chen, and M. Scanlon, IPv6 Security and Forensics, {\em In Proceedings of Sixth International Conference on Innovative Computing Technology (INTECH)}, pp. 743-748, IEEE, 2016.

\bibitem{vsl:Google} Google IPv6 Adoption Statistics (\url{https://www.google.com/intl/en/ipv6/statistics.html}), [Accessed January 2018].

\bibitem{vsl:Yasinsac} A. Yasinsac, and M. Yanet, Policies to enhance computer and network forensics. {\em In Proceedings of the 2001 IEEE Workshop on Information Assurance and Security}, pp. 289-295, 2001.

\bibitem{vsl:Nelson} B. Nelson, A. Phillips, and C. Steuart. {\em Guide to Computer Forensics and Investigations}, Cengage Learning, 2014.

\bibitem{vsl:bysync} M. Scanlon, J. Farina, and M-T. Kechadi, Network Investigation Methodology for BitTorrent Sync: A Peer-to-Peer Based File Synchronisation Service, {\em Computers \& Security}, vol. 54, pp. 27--43, 2015.

\bibitem{vsl:Walnycky} D. Walnycky, I. Baggili, A. Marrington, J. Moore, and F. Breitinger, Network and Device Forensic Analysis of Android Social-messaging Applications. {\em Digital Investigation}, vol. 14, pp. S77-S84, 2015.

\bibitem{vsl:hannaway} M. Scanlon, A. Hannaway, and M-T. Kechadi, A Week in the Life of the most popular BitTorrent swarms, {\em In The 5th Annual Symposium on Information Assurance (ASIA'10): Academic Track of 13th Annual NYS Cyber Security Conference}, Albany, New York, USA, June 2010.

\bibitem{vsl:Quick} D. Quick, and K-K. R. Choo, Impacts of Increasing Volume of Digital Forensic Data: A Survey and Future Research Challenges, {\em Digital Investigation}, vol. 11, no. 4, pp 273--294, 2014.

\bibitem{vsl:battling} M. Scanlon, Battling the Digital Forensic Backlog through Data Deduplication, {\em In Proceedings of 6th International Conference on Innovative Computing Technology (INTECH)}, pp. 10-14, IEEE, 2016.

\bibitem{vsl:Lillard} T. V. Lillard, {\em Digital Forensics for Network, Internet, and Cloud Computing: A Forensic Evidence Guide for Moving Targets and Data}. Syngress Publishing, 2010.

\bibitem{vsl:cloud} J. Farina, M. Scanlon, N-A. Le-Khac, and M-T Kechadi, Overview of the Forensic Investigation of Cloud Services, {\em In Proceedings of 10th International Conference on Availability, Reliability and Security (ARES)}, pp. 556-565, IEEE, 2015.

\bibitem{vsl:Casey} E. Casey, Network Traffic as a Source of Evidence: Tool Strengths, Weaknesses, and Future Needs, {\em Digital Investigation}, vol. 1, no. 1 pp. 28--43, 2004.

\bibitem{vsl:Hjelmvik} E. Hjelmvik, Passive Network Security Analysis with NetworkMiner, {\em (IN)SECURE Magazine}, no. 18, pp. 18--21, 2008.

\bibitem{vsl:Costa} G. Costa, and A.D. Franceschi, Xplico: Open Source Network Forensic Analysis Tool (NFAT) (\url{http://www.xplico.org/}), [Accessed January 2018].

\bibitem{vsl:Wireshark} Wireshark (\url{https://www.wireshark.org/}), [Accessed January 2018].

\bibitem{vsl:Zalewski} M. Zalewski, p0f: Passive OS Fingerprinting Tool (\url{http://lcamtuf.coredump.cx/p0f3/}, [Accessed January 2018].

\bibitem{vsl:Justniffer} Justniffer, (\url{http://justniffer.sourceforge.net/}), [Accessed January 2018].

\bibitem{vsl:WinPcap} WinPcap, (\url{https://www.winpcap.org/}), [Accessed January 2018].

\bibitem{vsl:Olifer} N. Olifer and V. Olifer. {\em Computer Networks: Principles, Technologies and Protocols for Network Design}, Wiley Publishing, 2005.

\bibitem{vsl:Reassembly} {\em IP Datagram Reassembly Algorithms}, (\url{https://tools.ietf.org/html/rfc815}), [Accessed January 2018].

\bibitem{vsl:IPTraf} IPTraf, (\url{http://iptraf.seul.org/}), [Accessed January 2018].

\bibitem{vsl:Wagener} G. Wagener, A. Dulaunoy, and T. Engel, Towards an estimation of the accuracy of TCP reassembly in network forensics, {\em In Proceedings of 2nd International Conference on Future Generation Communication and Networking (FGCN)}, pp. 273-278, IEEE, 2008.

\bibitem{vsl:tcpick} tcpick - a TCP stream sniffer, tracker and capturer, (\url{http://tcpick.sourceforge.net/}), [Accessed January 2018].

\bibitem{vsl:Pelles} Pelles C - a Complete Development Kit for Windows, (\url{http://www.smorgasbordet.com/pellesc/}), [Accessed January 2018].

\bibitem{vsl:Spooren} J. Spooren, D. Preuveneers, and W. Joosen, Mobile Device Fingerprinting Considered Harmful for Risk-based Authentication, {\em In Proceedings of the Eighth European Workshop on System Security}, ACM, 2015.

\end{chapthebibliography}
\end{document}